\documentclass{article}
\usepackage{amssymb}
\usepackage{amsmath}
\usepackage{amssymb}
\usepackage{amsmath}
\usepackage{amsthm}
\usepackage{bm,bbm}
\usepackage{geometry}
\usepackage{graphicx}
\usepackage{caption}
\usepackage{enumerate}
\usepackage[numbers]{natbib}
\usepackage{algorithm}
\usepackage{algpseudocode}
\usepackage[colorlinks,CJKbookmarks=true,bookmarksnumbered,linkcolor=black,citecolor=black,plainpages=true,pdfstartview=FitH]{hyperref}
\theoremstyle{plain}
\newtheorem{theorem}{Theorem}

\newtheorem{corollary}[theorem]{Corollary}

\newtheorem{lemma}[theorem]{Lemma}

\newtheorem{reduction}[theorem]{Reduction Rule}
\theoremstyle{definition}
\newtheorem{definition}[theorem]{Definition}

\theoremstyle{remark}

\usepackage[utf8]{inputenc}

\newcommand{\true}{\mathsf{true}}
\newcommand{\false}{\mathsf{false}}
\newcommand{\mvc}{\mathsf{vc}}
\newcommand{\poly}{\operatorname{poly}}

\title{An Improved Algorithm for Quantified Boolean Formulas}
\author{Zhiyang Chen, Hailong Yao}
\date{}

\begin{document}

\maketitle
\begin{abstract}
Quantified Boolean formula (QBF), the quantified version of Boolean satisfiability (SAT), is a fundamental problem in constraint solving. Due to the $\mathsf{PSPACE}$-complete nature of QBF, designing theoretically efficient QBF algorithms is much harder and receives less attention in academia. In this work, we present a new algorithm based on memoized search, which solves QBF with $n$ variables and $m$ clauses in $1.685^m\cdot\poly(n)$ time and space. This is the first improvement over the $1.709^m\cdot\poly(n)$ bound by \citet{williams02algorithms} in over 20 years. Our technique is based on the minimum vertex cover of the conflict graphs, which was proposed to study the parameterized complexity of SAT. We extend this idea to QBF and further show that QBF is slicewise polynomial-time solvable using this parameter.
\end{abstract}

\section{Introduction}
Boolean satisfiability (SAT) is a fundamental computational problem both in theory and practice. From the theoretical perspective, SAT is the prototype problem for $\mathsf{NP}$. Therefore, a wide variety of work aims to characterize the computational hardness of $\mathsf{NP}$ problems by studying the time complexity of SAT.

Given a Boolean formula in the conjunctive normal form (CNF) with $n$ variables, it is conjectured that SAT cannot be solved in $O^*((2-\varepsilon)^n)$ time for any constant $\varepsilon>0$ (we use $O^*(\cdot)$ to hide polynomial factors). However, if the clause width is at most $k$ (a.k.a. $k$-SAT) for some constant $k\ge 3$, it can be solved in $O^*(2^{(1-\Omega(1/k))n})$ time simultaneously by the PPSZ algorithm~\cite{scheder17ppsz}, local search~\cite{schoning99probabilistic}, and the polynomial method~\cite{chan21deterministic}. The exponential-time hypothesis (ETH)~\cite{impagliazzo01complexity} conjectures that 3-SAT cannot be solved in $2^{o(n)}$ time. Recently, \citet{openai26exponential} refutes ETH by presenting an algorithm that solves $k$-SAT in $2^{O_k(n\log\log n/\sqrt{\log n})}$ time using the polynomial method. Although this also refutes SETH (i.e., for any constant $\varepsilon>0$, there exists sufficiently large $k$ such that $k$-SAT cannot be solved in $O^*((2-\varepsilon)^n)$ time), their exponent is dependent on the clause width $k$. Therefore, it does not work for general CNF-SAT where the clauses can be arbitrarily wide. Another important parameterization is $m$, the number of clauses. A series of work~\cite{hirsch98two,yamamoto05improved,chu21improved} improves the complexity bound to $O^*(1.223^m)$. This breaks the trivial $O^*(2^n)$ bound on sparse SAT instances with $m\le cn$ for some constant $c\ge 1$. In fact, \citet{calabro06duality} show that there is a duality between $k$-SAT and CNF-SAT with a constant clause-to-variable ratio $m/n$. The refutation of ETH~\cite{openai26exponential} also implies that any SAT instances with $m=O(n)$ (or slightly super-linear) clauses can be solved in $O^*((2-\varepsilon)^n)$ time for some $\varepsilon>0$.

Quantified Boolean formula (QBF), a.k.a., quantified SAT (QSAT), is a natural extension of SAT by allowing arbitrary alterations of existential and universal quantifiers on Boolean variables. Though computationally harder, QBF has a variety of applications in practice, including formal verification, circuit synthesis, and games~\cite{shukla19survey}. It is also the canonical $\mathsf{PSPACE}$-complete analog to the $\mathsf{NP}$-complete SAT problem. Therefore, designing efficient QBF algorithms is both of practical and theoretical value.

However, QBF receives less research attention due to its hardness. Given a QBF instance with $n$ variables and $m$ clauses, even if the clause width is at most 3 (i.e., quantified 3-SAT), designing an $O^*((2-\varepsilon)^n)$-time algorithm is at least as hard as refuting SETH~\cite{calabro13on}. In fact, due to the $\mathsf{PSPACE}$-completeness, we believe quantified 3-SAT is much harder than $k$-SAT for any $k$. With respect to the clause number parameter $m$, the only non-trivial algorithm is due to \citet{williams02algorithms}, which solves QBF in $O^*(1.709^m)$ time. This bound can be improved to $O^*(1.619^m)$ for quantified 3-SAT.

\subsection{Our contributions}
In this work, we improve the time complexity for QBF to $O^*(1.685^m)$. To our knowledge, this is the first improvement to \citet{williams02algorithms} in over 20 years.
\begin{theorem}\label{thm:main}
    There exists an algorithm that returns the satisfiability of any QBF in deterministic $O^*(1.685^m)$ time and space.
\end{theorem}

The proposed algorithm is a memoized search, which is methodologically similar to \citet{williams02algorithms}. However, we perform a refined analysis on this algorithm. \citet{ganian21new} propose a new structure for SAT formulas, namely the conflict graph, and study the complexity of SAT using its structural parameters. Motivated by this work, we show that our QBF algorithm's complexity is closely related to the size of the minimum vertex cover (MVC) of the conflict graph. Formally, we prove two combinatorial bounds on the number of reachable states of the search algorithm. Such bounds are turned into the numerical $O^*(1.685^m)$ bound using computer-aided convex optimization techniques.

Moreover, we find that if the global conflict graph of the formula has a small MVC, we can further improve the time complexity. For example, if the MVC is at most $0.1m$, our bound can be improved to $O^*(1.369^m)$. This motivates us to study the MVC of the conflict graph as an algorithmic parameter for QBF. We show that, using this parameter, QBF is $\mathsf{W}[1]$-hard but slicewise polynomial-time solvable ($\mathsf{XP}$). However, if the clause width is upper-bounded by a constant $k$ (i.e., quantified $k$-SAT), this parameterization leads to a fixed-parameter tractable ($\mathsf{FPT}$) algorithm.

\subsection{Related Work}
\noindent\textbf{Algorithms for QBF}. As previously discussed, the only complexity upper bound parameterized by the number of clauses is due to \citet{williams02algorithms}. There are also a few research studies working on QBF algorithms from other perspectives. \citet{Santhanam15beating} show that QBF with few or many quantifier blocks can be solved faster than exhaustive search. Concretely, the hardest QBF instances are those where the number of quantifier blocks lies between $\Omega(\log n/\log\log n)$ and $O(\log n)$. \citet{Eriksson24solving} develop fixed-parameter tractable algorithms for quantified $k$-SAT parameterized by the number of existential variables, together with corresponding hardness results.

\noindent\textbf{Parameterized complexity of SAT}. Parameterized complexity research on SAT has developed several approaches to exploiting the hidden structure of CNF formulas. Backdoor sets~\cite{williams03backdoors} capture distance to tractable SAT classes. Structural parameters, particularly the treewidth of the primal, incidence, or related graphs, enable fixed-parameter tractable algorithms~\cite{fomin15dsat,ganian17backdoor}. \citet{ganian21new} investigate the treewidth of clause-based consensus and conflict graphs, establishing fixed-parameter tractability for model counting under the treewidth of consensus graphs while showing that SAT is $\mathsf{W}[1]$-hard parameterized by the treewidth of conflict graphs. Our work is motivated by their parameterized complexity results on conflict graph structures.

\subsection{Organization}
The rest of this paper is organized as follows: Section~\ref{sec:preliminaries} introduces the notation and the detailed problem formulation of this work. Section~\ref{sec:algorithm} presents our main result --- our memoized search algorithm and the time complexity bound. Section~\ref{sec:parameter} further studies the parameterized complexity using the minimum vertex cover of the conflict graph. Section~\ref{sec:conclusion} concludes the paper.

\section{Preliminaries}\label{sec:preliminaries}
\noindent\textbf{Notation}. We define $[n]=\{1,2,\dots,n\}$ for any integer $n\ge 1$. We use $\binom{a}{b}=\frac{a!}{b!(a-b)!}$ to denote binomial coefficients. If the parameters $a,b$ are invalid (e.g., $a>b$ or $a<0$), we set $\binom{a}{b}=0$. Moreover, we use $$h(x)=-x\log_2 x-(1-x)\log_2(1-x),$$ $$h(0)=h(1)=0$$ to denote the {\em binary entropy function} ($x\in [0,1]$).

\noindent\textbf{Problem formulation}. A Boolean formula on $n$ variables $x_1,\dots,x_n$ in the {\em conjunctive normal form} (CNF) is defined by $\mathcal{F}=C_1\wedge C_2\wedge\dots\wedge C_m,$ where each $C_i$ is a clause in the form $C_i=l_{i,1}\vee l_{i,2}\vee\dots\vee l_{i,k_i}$, and each literal $l_{ij}$ is a variable $x_q$ or its negation $\neg x_q$ for some $q\in[n]$. We study the standard prenex CNF form of quantified Boolean formulae, i.e., $$(Q_1 x_1) (Q_2 x_2)\dots(Q_n x_n)\, \mathcal{F},\ Q\in\{\forall,\exists\}^n,$$ where $\mathcal{F}$ is a CNF. Throughout this paper, we use $n$ to denote the number of variables, and $m$ to denote the number of clauses. The {\em quantified Boolean formula} (QBF) problem is to determine the satisfiability of a prenex CNF formula.

For simplicity of notation, we may treat a CNF as a set of clauses and treat a clause as a set of literals, i.e., $\mathcal{F}=\{C_1,\dots,C_m\}$ and $C_i=\{l_{i,1},\dots,l_{i,k_i}\}$. For any $i\in \{0,1,\dots,n\}$ and $R\subseteq [m]$, we define $$\mathcal{F}_i(R)=\{C_j\setminus\{\text{literals on}\ x_1,\dots,x_i\}\,|\,j\in R\}.$$

In the following, we recall several fundamental reduction rules in \citet{williams02algorithms}.
\begin{reduction}\label{rule:subproblem}
    Suppose the values of the first $i$ variables $x_1,\dots,x_i$ have been assigned, and $\mathcal{F}_i\subseteq \mathcal{F}$ is the set of unsatisfied clauses. We can safely remove all literals on $x_1,\dots,x_i$ in $\mathcal{F}_i$.
\end{reduction}
This reduction rule is obvious, since if a literal on $x_1,x_2,\dots, x_i$ appears in an unsatisfied clause, it must be unsatisfied and has no contribution to the clause. Therefore, it suffices to consider $\mathcal{F}_i(R)$ after assigning the values of the first $i$ variables. This allows us to apply dynamic programming (as in \citet{williams02algorithms}) or memoized search for QBF.

\begin{reduction}\label{rule:monotone}
    If a variable $x_i$ with quantifier $\exists$ (resp. $\forall$) occurs only positively or negatively in the formula, we can always set its literals to be satisfied (resp. unsatisfied).
\end{reduction}
This rule is known as the monotone literal rule. If the literals of a variable are all the same, we do not need to branch on this variable.

\begin{reduction}\label{rule:base}
    If there exists a clause $C_i$ such that all literals are false, the formula is unsatisfiable. If the formula has no clauses, the formula is satisfiable.
\end{reduction}

\section{An Improved QBF Algorithm}\label{sec:algorithm}
\subsection{Algorithm description}
The $O^*(1.709^m)$-time algorithm by \citet{williams02algorithms} is presented in the form of bidirectional search. Our improved algorithm is, in principle, similar to \citet{williams02algorithms}. However, for simplicity of analysis, we present it in the form of a memoized search algorithm. The algorithm is formally illustrated in Algorithm~\ref{alg:memo}, and $\textsc{MemoQBF}(0,\varnothing)$ returns the answer.

The state of our memoized search is in the form of $(i,R)$, which denotes that the first $i$ variables $x_1,\dots,x_i$ have been assigned and clauses in $R$ are still unsatisfied, i.e., we need to determine the satisfiability of $$Q_{i+1}x_{i+1}\cdots Q_nx_n\,\mathcal{F}_i(R).$$ By Reduction Rule~\ref{rule:subproblem}, this state is well-defined.

For each state, the algorithm branches on the next variable $x_{i+1}$. If $x_{i+1}$ is monotone, apply Reduction Rule~\ref{rule:monotone}. Otherwise, recursively search the two cases $x_{i+1}=\true$ and $x_{i+1}=\false$, and obtain the truth value based on the quantifier $Q_{i+1}$. The correctness of this algorithm is obvious. To analyze its complexity, the key is to bound the number of reachable states $(i,R)$ in the memoized search process. In the following subsections, we prove the $O^*(1.685^m)$ bound.
\begin{algorithm}[H]
\footnotesize
\caption{\textsc{MemoQBF}$(i,R)$}
\label{alg:memo}
\begin{algorithmic}[1]
\Require Prefix position $i$, unsatisfied clause set $R\subseteq[m]$;
\Ensure The truth value of $Q_{i+1}x_{i+1}\cdots Q_nx_n\,\mathcal{F}_i(R)$;
\If{$(i,R)$ is in the memo table}
   \State \Return the stored value;
\EndIf
\If{$R=\varnothing$}
   \State Store $\true$ and \Return $\true$; (Reduction Rule~\ref{rule:base})
\EndIf
\If{some clause of $\mathcal{F}_i(R)$ is empty}
   \State Store $\false$ and \Return $\false$; (Reduction Rule~\ref{rule:base})
\EndIf
\If{\(i=n\)}
   \State Store the value of $\mathcal{F}_i(R)$ and \Return it;
\EndIf
\State $P\gets \{j\,|\,j\in R,\,x_{i+1}\in C_j\}$;
\State $N\gets \{j\,|\,j\in R,\,\neg x_{i+1}\in C_j\}$;
\If{$P=N=\varnothing$}
   \State $v\gets\textsc{MemoQBF}(i+1,R)$;
\ElsIf{$P=\varnothing$ or $N=\varnothing$}
   \State Let $L$ be the nonempty one of $P,N$;
   \If{$Q_{i+1}=\exists$}
      \State $v\gets\textsc{MemoQBF}(i+1,R\setminus L)$; (Reduction Rule~\ref{rule:monotone})
   \Else
      \State $v\gets\textsc{MemoQBF}(i+1,R)$; (Reduction Rule~\ref{rule:monotone})
   \EndIf
\Else
   \State $v_0\gets\textsc{MemoQBF}(i+1,R\setminus N)$;
   \State $v_1\gets\textsc{MemoQBF}(i+1,R\setminus P)$;
   \If{$Q_{i+1}=\exists$}
      \State $v\gets v_0\vee v_1$;
   \Else
      \State $v\gets v_0\wedge v_1$;
   \EndIf
\EndIf
\State Store $v$ in the memo table;
\State \Return $v$;
\end{algorithmic}
\end{algorithm}

\subsection{Prefix conflict graphs}
Throughout our analysis, we fix the value of $i\in\{0,1,\dots,n\}$, and consider the number of states $R$ for this $i$. We define the prefix conflict graph $G_i$ for each $i$.
\begin{definition}
    The {\em prefix conflict graph} $G_i$ is an undirected graph with $m$ vertices, each representing a clause in $\mathcal{F}$. For any two clauses $a$ and $b$, an edge $(a,b)\in G_i$ exists if for some $1\le j\le i$, one clause contains the literal $x_j$ and the other one contains $\neg x_j$.
\end{definition}
In other words, the clauses $a$ and $b$ have a conflict on the variable $x_j$ and cannot be satisfied simultaneously by $x_j$.

Since we fix the value of $i$, in the following, we omit the subscript $i$ and use $G=G_i$ to denote the conflict graph. Let $U\subseteq [m]$ be a minimum vertex cover of $G$ and $\bar{U}=[m]\setminus U$ be a maximum independent set. (Recall that the minimum vertex cover (MVC) is the minimum subset of vertices such that every edge has at least one terminal in the MVC. The complement of MVC is the maximum independent set.) If multiple MVCs exist, we break ties arbitrarily. We use $\mvc=|U|$ to denote the size of the MVC.

For any reachable state $(i,R)$, let $D=[m]\setminus R$ be the set of deleted clauses.
\begin{lemma}\label{lmm:D-cover}
    For any reachable state $(i,R)$, $R$ is an independent set in $G$. Therefore, $D$ is a vertex cover of $G$.
\end{lemma}
\begin{proof}
    If $a,b\in R$ since clause $a$ contains $x_j$ and $b$ contains $\neg x_j$ for some $j\le i$, at least one of them is satisfied by $x_j$ and cannot simultaneously remain in $R$.
\end{proof}
Bounding the complexity of Algorithm~\ref{alg:memo} essentially reduces to bounding the number of reachable states $(i,R)$. Note that in the memoized search, the search process is represented by a directed acyclic graph (DAG). We can ``unroll'' the DAG into a search tree by duplicating the nodes with the same state $(i,R)$.

Before diving into the proof of the main theorem, we intuitively discuss why the minimum vertex cover of the conflict graph helps tighten the bound.
\begin{itemize}
    \item On the one hand, at a search tree state that branches on the next variable, let $P$ and $N$ be the active positive and negative clause sets for the variable. For every pair $a\in P$ and $b\in N$, $(a,b)$ is an edge of the conflict graph $G$. Therefore, a vertex cover must include one complete side: $P\subseteq U$ or $N\subseteq U$. In the search tree, $P$ and $N$ correspond to two children of the current state (and two values possibly assigned to the next variable). We mark this complete child as an ``important'' state. Fix the value of $i$. If we determine which of the first $i$ variables $x_1,\dots,x_i$ have been assigned an important value, the current state is also determined. Therefore, this argument contributes to a good bound on the number of reachable states if $\mvc$ is small.
    \item On the other hand, the minimum vertex cover can be used to analyze the number of possible remaining variable sets $R$. Let $X=R\cap U$ be the vertices in the MVC that have not been satisfied. Note that $X$ is independent. Replacing $X$ inside $U$ by its neighbors outside $U$ must not produce a smaller vertex cover, because $U$ is an MVC. Therefore, $X$ has at least $|X|$ neighbors outside $R$. Those neighbors cannot survive together with $X$, so they belong to the deleted variable set $D$. This argument produces a good bound if $\mvc$ is large.
\end{itemize}

In the following analysis, we fix the MVC size $\mvc$, the number of deleted clauses $d=|D|$, the number of deleted clauses in the MVC $\mvc_D=|D\cap U|$, and bound the number of reachable states for specific values of $i,\mvc,d,\mvc_D$. Since all these parameters lie in $\{0,1,\dots,\max\{n,m\}\}$, the union of reachable states for all their possible values only contributes to a polynomial factor in the complexity.

\noindent\textbf{Counting branching paths}. Consider the conflict graph $G$. For every $j\le i$, let $P_j$ be the set of clauses containing $x_j$ and $N_j$ be the set of clauses containing $\neg x_j$. We have $P_j\subseteq U$ or $N_j\subseteq U$. (Otherwise, any clauses $a\in P_j$ and $b\in N_j$ forms an edge $(a,b)\in G$ with neither terminal in the vertex cover $U$, which leads to a contradiction.) Let $U_j$ be the one that lies in $U$. If both $P_j\subseteq U$ and $N_j\subseteq U$, we select one using an arbitrary tie-breaking rule. For any search tree node that branches on $x_j$, it has two children corresponding to $x_j=\true$ and $x_j=\false$, respectively. We mark the child that satisfies the clauses $U_j$ as ``important''. After fixing $i,\mvc,d,\mvc_D$ and the important values of $x_j$ for $j\in [i]$, the state $R$ is determined solely by whether $x_1,\dots,x_i$ are assigned important values.
\begin{lemma}\label{lmm:T1}
    For fixed $i,\mvc,d,\mvc_D$, the number of reachable states $(i,R)$ with $|D|=d$ and $|D\cap U|=\mvc_D$ is at most $$T_1=\sum_{l=0}^d\sum_{g=0}^{\min\{\mvc_D,l\}}\binom{l}{g}=\sum_{g=0}^{\mvc_D}\binom{d+1}{g+1}.$$
\end{lemma}
\begin{proof}
    Consider the path from the root $(0,\varnothing)$ to an arbitrary search tree node. In this path, some nodes are unary (lines 15--23 of Algorithm~\ref{alg:memo}) since the branched variable occurs only positively or negatively in the remaining formula (Reduction Rule~\ref{rule:monotone}). Other nodes are binary, each of which has one child selected in the path. For binary nodes, we mark them with 1 if the selected child is important and 0 otherwise. Suppose altogether there are $l$ binary nodes in the path. The path can be uniquely represented by a binary string of length $l$. Therefore, it suffices to bound the number of binary strings, where the number of ones is at most $\mvc_D$.

    Note that $l\le \min\{i,d\}$, since only the first $i$ variables are branched and each of the binary node deletes at least one clause (both $P$ and $N$ are non-empty). In this proof, we only use the inequality $l\le d$. We enumerate $l$ from 0 to $d$, and enumerate the number of ones in the binary string for each $l$. This leads to the bound $$T_1=\sum_{l=0}^d\sum_{g=0}^{\min\{\mvc_D,l\}}\binom{l}{g}.$$ The latter equality comes from the Hockey-Stick Identity $$\sum_{k=r}^n \binom{k}{r}=\binom{n+1}{r+1}.$$
\end{proof}

\noindent\textbf{Counting sets of remaining clauses}. The second part of our analysis directly counts the number of states $R$. In the following analysis, let $X=R\cap U$ be the vertices in the MVC that have not been satisfied, and $N_{\bar{U}}(X)=\{c\,|c\in \bar{U}, \,\exists\, c^\prime\in X\ \text{s.t.}\ (c,c^\prime)\in G\}$ be the neighbors of $X$ in $\bar{U}$. We have the following lemma.
\begin{lemma}\label{lmm:expansion}
    We have $$N_{\bar{U}}(X)\subseteq D\cap \bar{U},\ |N_{\bar{U}}(X)|\ge |X|=\mvc-\mvc_D.$$
\end{lemma}
\begin{proof}
    By Lemma~\ref{lmm:D-cover}, $R$ is an independent set of $G$, which means the neighbors of $X\subseteq R$ all lie in $[m]\setminus R=D$. Therefore, $N_{\bar{U}}(X)\subseteq D\cap \bar{U}$.

    Assume $|N_{\bar{U}}(X)|<|X|$. Since $X$ is an independent set, $$U^\prime=(U\setminus X)\cup N_{\bar{U}}(X)$$ is also a vertex cover of $G$. Then, $|U^\prime|=\mvc-|X|+|N_{\bar{U}}(X)|<\mvc$, which contradicts the minimality of $U$.
\end{proof}

Therefore, we can bound the number of states by the following result.
\begin{lemma}\label{lmm:T2}
    For fixed $i,\mvc,d,\mvc_D$, the number of reachable states $(i,R)$ with $|D|=d$ and $|D\cap U|=\mvc_D$ is at most $$T_2=\binom{\mvc}{\mvc_D}\binom{m-2\mvc+\mvc_D}{d-\mvc}.$$
\end{lemma}
\begin{proof}
    Consider the clauses in the MVC $U$. Since $U$ is fixed, the number of choices for $X=R\cap U$ among $U$ is $\binom{\mvc}{\mvc-\mvc_D}=\binom{\mvc}{\mvc_D}$ (recall that $\mvc_D=|D\cap U|$).
    
    Now, we fix $X$, and consider the number of choices for $R\setminus U=R\cap\bar{U}$. This also fixes $N_{\bar{U}(X)}$. Since $R=[m]\setminus D$, it suffices to consider the number of choices for $Y=D\setminus U=D\cap \bar{U}$, where $|Y|=d-\mvc_D$.
    
    The $N_{\bar{U}}(X)$ part in $Y$ is fixed. By Lemma~\ref{lmm:expansion}, $N_{\bar{U}}(X)\subseteq Y$, and $|N_{\bar{U}}(X)|\ge\mvc-\mvc_D$. Fix any $\mvc-\mvc_D$ of them. Any valid $Y$ contains them, leaving at most $(d-\mvc_D)-(\mvc-\mvc_D)=d-\mvc$ further vertices to choose from $|\bar{U}|-(\mvc-\mvc_D)=m-2\mvc+\mvc_D$ vertices. The total number of choices is thus at most $$\binom{m-2\mvc+\mvc_D}{d-\mvc}.$$ Note that the binomial is always valid: Since $D$ is a vertex cover (Lemma~\ref{lmm:D-cover}), we have $d\ge\mvc$ by the minimality of $U$. We obviously have $0\le \mvc_D\le \mvc$. Moreover, $D\cap \bar{U}\subseteq \bar{U}$ yields $d-\mvc_D\le m-\mvc$. Therefore, $0\le d-\mvc\le m-2\mvc+\mvc_D$.
    
    Finally, multiplying by the choices of $X$ yields the desired bound.
\end{proof}
Combining Lemmas~\ref{lmm:T1} and~\ref{lmm:T2}, we have the following theorem.
\begin{theorem}
    The time (and space) complexity of Algorithm~\ref{alg:memo} is bounded from above by $\min\{T_1,T_2\}\cdot\poly(n,m)$.
\end{theorem}

\subsection{A numerical bound}\label{sec:numerical}
In this subsection, we derive a numerical upper bound for $\min\{T_1,T_2\}$ to prove the main theorem. For simplicity of analysis, we normalize the parameters $d,\mvc,\mvc_D$. Let $$\alpha=\frac{d}{m},\,\beta=\frac{\mvc}{m},\,\gamma=\frac{\mvc_D}{m}.$$ The feasible domain for $(\alpha,\beta,\gamma)$ is the polyhedron $$
\mathcal{P}=
\left\{(\alpha,\beta,\gamma)\bigg|
\begin{array}{l}
 0\le\gamma\le\beta\le\alpha\le 1,\\
 \alpha+\beta-\gamma\le 1
\end{array}
\right\}.
$$ This polyhedron encodes the following inequalities: (1) $d\ge\mvc$ (Lemma~\ref{lmm:D-cover}); (2) $0\le\mvc_D\le\mvc$ ($D\cap U\subseteq U$); (3) $d-\mvc_D\le m-\mvc$ ($D\cap \bar{U}\subseteq \bar{U}$).

Recall that $h(x)$ is the binary entropy function. We define the single-sided entropy function $$\bar{h}(x)=
    \begin{cases}
        h(x),&0\le x\le\frac{1}{2},\\
        1,&\frac{1}{2}\le p\le 1.
    \end{cases}
$$ We have the following estimation of the binomial sum.
\begin{lemma}\label{lmm:binomial}
    For any $n\ge 1$ and $0\le k\le n$, we have $$\binom{n}{k}\le 2^{nh(k/n)},\ \ \sum_{i=0}^k\binom{n}{i}\le (n+1)2^{n\bar{h}(k/n)}.$$
\end{lemma}
\begin{proof}
    The binomial theorem gives $$1=(p+(1-p))^n\ge \binom{n}{k}p^k(1-p)^{n-k}.$$ Plugging in $p=k/n$ yields $$\binom{n}{k}\le 2^{nh(k/n)}.$$

    The second inequality has two cases: If $k\le n/2$, since $\binom{n}{i}$ is increasing for $i\le k$, we have $$\sum_{i=0}^k\binom{n}{i}\le (n+1)2^{nh(k/n)}.$$ If $k>n/2$, the sum is at most $2^n=2^{n\bar{h}(k/n)}$ since $\bar{h}(k/n)=1$.
\end{proof}
\begin{proof}[Proof of Theorem~\ref{thm:main}]
    Applying the binomial bounds in Lemma~\ref{lmm:binomial}, Lemmas~\ref{lmm:T1} and~\ref{lmm:T2} give $$T_1\le \poly(n,m)\cdot 2^{m\cdot P_1(\alpha,\gamma)},$$ $$T_2\le \poly(n,m)\cdot 2^{m\cdot P_2(\alpha,\beta,\gamma)},$$ where $$P_1(\alpha,\gamma)=\alpha\cdot\bar{h}\left(\frac{\gamma}{\alpha}\right),$$ and $$P_2(\alpha,\beta,\gamma)=\beta\cdot h\left(\frac{\gamma}{\beta}\right)+(1-2\beta+\gamma)\cdot h\left(\frac{\alpha-\beta}{1-2\beta+\gamma}\right).$$

    Therefore, it suffices to bound $$\max_{(\alpha,\beta,\gamma)\in\mathcal{P}}\min\{P_1(\alpha,\gamma),P_2(\alpha,\beta,\gamma)\}.$$ It is easy to verify that both $P_1$ and $P_2$ are concave functions over $\mathcal{P}$, and thus $\min\{P_1,P_2\}$ is also concave. We can compute the maximum of $\min\{P_1,P_2\}$ using standard numerical optimization algorithms, such as gradient ascent. A computer-aided proof show that $\min\{P_1(\alpha,\gamma),P_2(\alpha,\beta,\gamma)\}\le P^*=0.7521$, where the maximum is taken at $$\alpha^*\approx 0.7543,\,\beta^*\approx 0.4526,\,\gamma^*\approx 0.3531.$$ This yields $$\min\{T_1,T_2\}\le \poly(n,m)\cdot 2^{m\cdot\min\{P_1,P_2\}}\le \poly(n,m)\cdot 2^{m\cdot P^*}\le O^*(1.685^m)$$ as we desire.
\end{proof}

\section{Parameterized complexity by conflict graph MVC}\label{sec:parameter}
Recall that in the proof of our main theorem, we omit the subscript $i$ for the prefix conflict graph $G_i$. Obviously, we have $G_1\subseteq G_2\subseteq\dots\subseteq G_n$, and $G_n$ is the global conflict graph that considers all $n$ variables. Therefore, let $\mvc_i$ be the size of the MVC for $G_i$. We have $\mvc_1\le \mvc_2\le \dots\le\mvc_n$. The proof indicates that if the minimum vertex cover of the global conflict graph $\mvc_n$ is small (below $0.4526m$), we can obtain stronger complexity bounds for Algorithm~\ref{alg:memo}.

Concretely, if $\mvc_n\le c\cdot m$, we can compute the complexity bound by $$P^*=\max_{\substack{(\alpha,\beta,\gamma)\in\mathcal P\\ \beta\le c}}\min\{P_1(\alpha,\gamma),P_2(\alpha,\beta,\gamma)\}.$$ We list some numerical examples in Table~\ref{tab:smallmvc}. If the minimum vertex cover of the global conflict graph is small, we can solve QBF faster. It is hence natural to study further the computational complexity of QBF parameterized by the MVC.
\begin{table}[ht]
    \centering
    \begin{tabular}{c|cc|c}\hline
        $\mvc$ & $\alpha^*$ & $\gamma^*$ & Complexity bound \\\hline
        $0.1m$ & 0.9 & 0.1 & $1.369^m$ \\
        $0.2m$ & 0.8030 & 0.1994 & $1.569^m$ \\
        $0.3m$ & 0.7708 & 0.2738 & $1.652^m$ \\
        $0.4m$ & 0.7595 & 0.3278 & $1.681^m$ \\
        $\ge 0.4526m$ & 0.7543 & 0.3531 & $1.685^m$ \\\hline
    \end{tabular}
    \caption{Examples of complexity bounds for small MVCs.}
    \label{tab:smallmvc}
\end{table}

We first recall some basic concepts of parameterized complexity. For any computation problem, let $n$ denote the input length and $k$ be a parameter of the input. We say this problem is {\em fixed-parameter tractable} ($\mathsf{FPT}$) if it can be solved in $f(k)\cdot\poly(n)$ time for some computable function $f$. The problem is {\em slicewise polynomial} solvable ($\mathsf{XP}$) if it can be solved in $O(n^{f(k)})$ time for some computable function $f$.

Is QBF parameterized by the MVC of the conflict graph in $\mathsf{FPT}$? \citet{ganian21new} shows that the answer is negative even for SAT assuming $\mathsf{W}[1]\ne\mathsf{FPT}$.
\begin{theorem}[Proposition 1 of \citet{ganian21new}]
    SAT is $\mathsf{W}[1]$-hard when parameterized by the size of the minimum vertex cover of the conflict graph.
\end{theorem}

However, this problem is indeed in $\mathsf{XP}$. This result can be proved using Algorithm~\ref{alg:memo} even without memoization.
\begin{theorem}\label{thm:xp}
    Let $\mvc$ denote the size of the MVC for the global conflict graph of a QBF instance. QBF can be solved in $O^*(k^{\mvc})$ time and $\poly(n,m)$ space, where $3\le k\le n$ is the maximum width of clauses.
\end{theorem}
\begin{proof}
    We consider Algorithm~\ref{alg:memo} without storing the truth values in the memo table. Since it is a depth-first search algorithm, only polynomial space is used. It suffices to consider the time complexity. Besides Reduction Rules~\ref{rule:subproblem} to~\ref{rule:base}, we have another important rule, the {\em unit propagation}.
    \begin{reduction}
        If a clause has only one literal, set its variable to satisfy the literal when the quantifier is existential. If the quantifier is universal, return false.
    \end{reduction}
    
    Between lines 12 and 13 of Algorithm~\ref{alg:memo}, we apply this reduction rule exhaustively to ensure that all remaining clauses have width at least 2. Now, let $w(C)$ denote the width of any clause $C$. Fix an MVC $U$ for the conflict graph. We define a potential function $$\Phi=\prod_{\text{live clause }C\in U} w(C),$$ which is the product of widths for all remaining clauses in $R$. If $R$ is empty, the product is 1. Since all reduction rules and branches only remove clauses or literals, $\Phi$ is non-increasing with respect to the search depth.

    At a binary branch on variable $x$, with $P$ and $N$ defined in the algorithm, recall that $U$ must include one complete side, $P\subseteq U$ or $N\subseteq U$. Choose one clause $C$ on such a side with $w(C)\ge 2$. In the child satisfying this side, $C$ is satisfied, and the $w(c)$ factor is removed. In the other child, $C$ loses a literal. Let $\Phi_{\text{satisfy}}$ and $\Phi_{\text{falsify}}$ be the potential functions for two children. We have $$\Phi_{\text{satisfy}}\le\frac{\Phi}{w(C)},\,\Phi_{\text{falsify}}\le\frac{(w(C)-1)\Phi}{w(C)},$$ which implies $$\Phi_{\text{satisfy}}+\Phi_{\text{falsify}}\le \Phi.$$ Therefore, we have (1) search tree leaves have potential at least 1; (2) unary nodes do not increase the potential; (3) the sum of potentials for two children of a binary node does not exceed the parent. An induction on the search tree easily shows that the number of leaves does not exceed the potential of the root, i.e., $\Phi\le \prod_{C\in U} w(C)\le k^\mvc$.
\end{proof}
\begin{corollary}
    QBF parameterized by the MVC of the conflict graph is in $\mathsf{XP}$. Moreover, quantified $k$-SAT parameterized by the MVC of the conflict graph is in $\mathsf{FPT}$ for any constant $k\ge 3$.
\end{corollary}

For quantified 3-SAT, we achieve an $O^*(3^{\mvc})$ time complexity. As a comparison, \citet{williams02algorithms} also studies the time complexity for quantified 3-SAT, which shows that as the clause-to-variable ratio varies between 2 and 1, the complexity bound grows from $O^*(1.415^m)$ to $O^*(1.619^m)$. Our result outperforms the $O^*(1.415^m)$ bound when $\mvc\le 0.3153m$.

\section{Conclusion}\label{sec:conclusion}
In this paper, we study the worst-case complexity of quantified Boolean formulas. For QBF with $m$ clauses, we present an algorithm in $O^*(1.685^m)$ time. The key to the complexity analysis is to exploit the minimum vertex cover of the conflict graph structures. We further study this minimum vertex cover as an algorithmic parameter and show that QBF is slicewise polynomial-time solvable under this parameterization.

\section*{Disclosure of LLM Usage}
The main result of this paper is motivated by discussions with GPT-5.6 Sol, who suggests using conflict graphs to study QBF algorithms. The computer-aided proof of the numerical bound in Section~\ref{sec:numerical} is derived by GPT-6 Sol. This paper is entirely written by human authors, and the authors take full responsibility for the content of this paper.

\bibliographystyle{unsrtnat}
\bibliography{edabib}

\appendix

\end{document}